\documentclass{article}

\usepackage[final,nonatbib]{neurips_2020}

\usepackage[utf8]{inputenc} 
\usepackage[T1]{fontenc}    
\usepackage{url}
\usepackage[final]{hyperref}       
\usepackage{booktabs}       
\usepackage{amsfonts}       
\usepackage{nicefrac}       
\usepackage{microtype}      

\usepackage[backend=biber, style=numeric, sorting=none]{biblatex}
\usepackage[final]{graphicx}

\title{Latent Space Oddity: Exploring Latent Spaces to Design Guitar Timbres}

\author{%
  Jason Taylor \\
    Montreal, Canada \\
  \texttt{jasonrbtaylor@gmail.com} \\
}

\begin{document}

\maketitle

\begin{abstract}
   We introduce a novel convolutional network architecture with an interpretable latent space for modeling guitar amplifiers. Leveraging domain knowledge of popular amplifiers spanning a range of styles, the proposed system intuitively combines or subtracts characteristics of different amplifiers, allowing musicians to design entirely new guitar timbres.
\end{abstract}

\section{Introduction}
Guitar amplifiers are responsible for significant signal modification, such as equalization (EQ), compression and distortion, to the extent that the resulting timbre is determined more by the amplifier than the guitar itself. Amplifiers utilizing vacuum tubes and analog circuitry are the most popular approach to creating guitar tones. However, these amplifiers have several drawbacks: cost, unreliability due to vacuum tube failure, and the dependence of timbre on the amplification level, leading to difficulties in achieving specific sounds at a required volume. Amplifiers are also limited in their timbral range, as they utilize only a few low-order analog filters for equalization (EQ) and typically 1 to 4 distortion stages. While digital amplifier simulators have gained mainstream success recently, these typically recreate the timbres of vacuum tube amplifiers by simulating the analog circuitry and are similarly limited in their timbral range \cite{fractal}. Digital amplifier simulators also suffer from a limited interface that mimics analog amplifiers, limiting users to adjusting coarse parameters like 3-band EQ and distortion levels.

This work enables musicians to create new guitar timbres by exploring the latent space of a convolutional network that was trained to reproduce a range of guitar amplifiers. We introduce a novel model architecture with an interpretable latent space. Interpolating parameters allows musicians to combine characteristics from various amplifiers, while extrapolating beyond the latent space spanned by the amplifiers in the training set generates entirely new timbres.


\section{Method and Experiments}
\begin{figure*}[t]
\centering
\includegraphics[width=\textwidth]{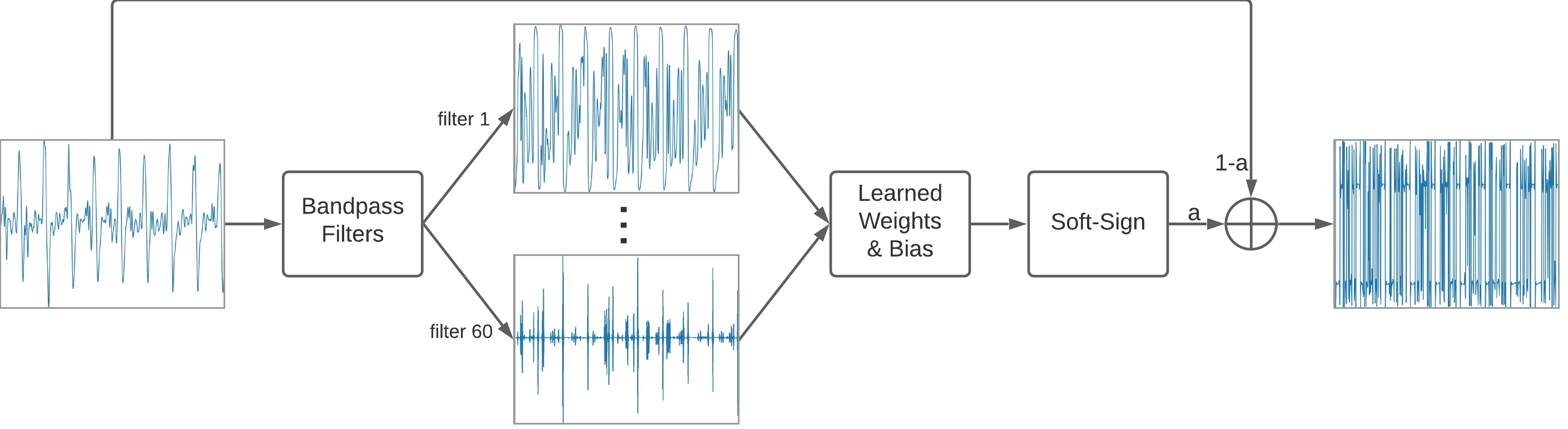}
\center \caption{Depiction of one EQ block with a fixed bandpass filter bank. The proposed model consists of six EQ blocks connected in series.}
\label{fig:EQlayer}
\end{figure*}

\paragraph{Architecture}
The model consists of six sequential EQ blocks shown in Figure \ref{fig:EQlayer}: convolution with 60 fixed bandpass filters, followed by a learned linear combination (reducing the dimensionality from 60 to 1), a learned bias to allow for asymmetric distortion, and a soft-sign activation function for waveform compression or distortion. Each EQ block has a residual connection \cite{resnet} with a learned parameter to control a convex combination of its input and output, to produce cleaner timbres as needed by skipping the soft-sign activation. The learned EQ weights are forced to be positive by squaring their values, which avoids phase cancellation in the latent features.

\paragraph{Filters}
Using fixed filters rather than learned convolutional weights aligns the latent spaces of all the models. This enables an intuitive latent space exploration for timbre design.

The filters cover a range of 40 Hz (approximately one octave below the lowest note on a guitar in standard tuning) to 20 kHz (the upper limit of human hearing). The bandpass center frequencies are logarithmically-spaced to follow the human perception of pitch \cite{pitch}. Finite-impulse response (FIR) filters, equivalent to 1D convolutions, are used due to their linear phase property to avoid phase distortion. Infinite impulse response (IIR) filters achieve sharper transitions around the bandpass frequencies and were explored for this work. However, IIR filters, analogous to recurrent networks, increase the training time by several orders of magnitude over FIR filters due to the length of the training inputs. The lower limit for perceptible latency in audio is approximately 2 milliseconds \cite{latency}, which serves as an upper bound for the total filter group delay across all layers in the model. 

\paragraph{Loss functions}
Mean-squared error losses in both the time domain and mel-frequency spectrogram domain \cite{mel} were explored. Qualitatively, models trained with time-domain losses, as in \cite{timeloss2, timeloss}, reproduce the distortion of the amplifiers but lack balance between treble and bass. Conversely, models trained with a mean-squared error loss in the mel-frequency spectrogram domain produce a more balanced sounding output. The sum of losses in both domains generalizes well across all amplifiers tested and is used in all results.

\paragraph{Data}
The training data consists of 50 minutes of direct-input (DI) signals. 
An additional 3 minutes of validation data is used to control early stopping. The following amplifiers are modeled (ordered from least distortion to most distortion): Fender Twin, Vox AC30, Marshall Super Lead, Marshall JCM800, Mesa Boogie Triple Rectifier, and Peavey 5150. Further details are in the supplement.

\section{Results and Discussion}

Model outputs for all six amplifiers with a test DI signal can be heard at \url{http://youtu.be/cbXa_N1sbvk} along with a visualization of the weights and network activations.

\paragraph{Latent Space Exploration}
We linearly interpolate between the model parameters of two amplifiers $\theta_i$ and $\theta_j$ as
\begin{equation} \label{eq:1}
\theta_{ij\alpha} = (1-\alpha)\, \theta_i + \alpha\, \theta_j
\end{equation}
where $\alpha$ $\in$ [0, 1] controls the blending of the characteristics of the two amplifiers. The same formulation (\ref{eq:1}) is used for extrapolation, with $\alpha$ $>$ 1 resulting in a model similar to amplifier $j$ that has fewer characteristics of amplifier $i$. To preserve the respective distortion levels of the amplifiers, the resulting parameters are not normalized. A user-controlled gain parameter can account for this, similar to a gain control on a guitar amplifier that adjusts the distortion. Note that the interpolation and extrapolation are not generally limited to combinations of two models but is done here for interpretability of the results.

The results of interpolation for $\alpha = 0.5$ between all pairs of amplifiers can be heard at \url{http://youtu.be/FYEN8pMdOnw}. Extrapolation results can be heard at \url{http://youtu.be/q-WrRKHQbBs} for $\alpha$ $\in$ $\{1.25, 1.5\}$. Interpolation for all amplifier pairs yields novel timbres combining characteristics of each amplifier that are not possible with previous amplifier modeling approaches. Extrapolation generates more original timbres for the modern guitarist and the examples provided cover only a small subset of what is possible with this system.

\begin{center}
    \printbibliography
\end{center}

\section*{Supplementary Materials}

\paragraph{Recording setup} All labels were recorded at 44.1 kHz. Due to equipment availability, a digital amplifier simulator (Line 6 Helix) was used to record the ground truth labels rather than recording the original amplifiers. A similar approach is used in \cite{timeloss}. Although this may propagate modeling errors, qualitatively these are not perceptibly significant and exact replication of the amplifier timbres is secondary to the goal of intuitive latent space exploration in this work.

\paragraph{Speaker model} The recorded sound of an amplifier includes the speakers and microphone, each of which impart changes to the tone. In this work we model the amplifier without the speaker and microphone. The combined speaker and microphone EQ response is added to the final result by convolution with a commercially-available impulse response (IR) file. A similar recording setup is commonly used for vacuum tube amplifiers with a USB load-box that emulates the electrical load of a speaker. Separating the amplifier and speaker models gives musicians greater control over the resulting timbre.

The speaker IR used for the results in this work was from an Orange 2x12 speaker cabinet with Celestion V30 and G12M-25 speakers recorded with a proprietary mix of microphones by OwnHammer. The same IR was used for all results to avoid conflating the variations in the models with different speakers.

\paragraph{DI files} The DI files used for training inputs were scraped from several guitar forums and concatenated temporally. Silences longer than 1 second were removed to avoid batches that did not contribute to the training. The validation and test DI were recorded by the author with different guitars. This setup avoids overfitting the model to the EQ profile of a particular guitar.

\paragraph{Societal Impact} This work does not present any foreseeable societal consequence.

\end{document}